\documentclass{ws-jai}
\usepackage[flushleft]{threeparttable}

\newcommand\blfootnote[1]{%
  \begingroup
  \renewcommand\thefootnote{}\footnote{#1}%
  \addtocounter{footnote}{-1}%
  \endgroup
}

\begin{document}

\catchline{}{}{}{}{} 

\markboth{N. Galitzki et al.}{The Next Generation BLAST Experiment}

\newcommand{\arcsec}{$^{\prime\prime}$}
\newcommand{\arcmin}{$^{\prime}$}
\newcommand{\micron}{$\,\mu$m}

\title{THE NEXT GENERATION BLAST EXPERIMENT}

\author{Nicholas Galitzki$^1$, Peter A. R. Ade$^2$, Francesco E. Angil{\`e}$^1$, Peter Ashton$^3$, James A. Beall$^4$, Dan Becker$^4$, Kristi J. Bradford$^5$, George Che$^5$, Hsiao-Mei Cho$^6$, Mark J. Devlin$^1$, Bradley J. Dober$^1$, Laura M. Fissel$^3$, Yasuo Fukui$^7$, Jiansong Gao$^4$, Christopher E. Groppi$^5$, Seth Hillbrand$^8$, Gene C. Hilton$^4$, Johannes Hubmayr$^4$, Kent D. Irwin$^9$, Jeffrey Klein$^1$, Jeff Van Lanen$^4$, Dale Li$^4$, Zhi-Yun Li$^{10}$, Nathan P. Lourie$^1$, Hamdi Mani$^5$, Peter G. Martin$^{11}$, Philip Mauskopf$^{2,5}$, Fumitaka Nakamura$^{12}$, Giles Novak$^3$, David P. Pappas$^4$, Enzo Pascale$^2$, Giampaolo Pisano$^2$, Fabio P. Santos$^3$, Giorgio Savini$^{13}$, Douglas Scott$^{14}$, Sara Stanchfield$^1$, Carole Tucker$^2$, Joel N. Ullom$^4$, Matthew Underhill$^5$, Michael R. Vissers$^4$, Derek Ward-Thompson$^{15}$}

\address{
$^1$Department of Physics and Astronomy University of Pennsylvania, 209 South 33rd Street, Philadelphia, PA 19104; \\
$^2$Department of Physics $\&$ Astronomy, Cardiff University, 5 The Parade, Cardiff CF24 3A A, UK;\\
$^3$Department of Physics $\&$ Astronomy, Northwestern University, 2145 Sheridan Road, Evanston, IL 60208; \\
$^4$Quantum Electronics and Photonics Division, National Institute of Standards and Technology, 325 Broadway Street, Boulder, CO 80305; \\
$^5$School of Earth and Space Exploration, Arizona State University, P.O. Box 876004, Tempe, AZ 85287; \\
$^6$SLAC National Accelerator Laboratory, 2575 Sand Hill Road, Menlo Park, CA 94025;\\
$^7$Department of Physics and Astrophysics, Nagoya University, Chikusa-ku Nagoya 464-8602, Japan; \\
$^8$Department of Physics and Astronomy, California State University, Sacramento, 6000 J Street, Sacramento, CA 95819; \\
$^9$Department of Physics, Stanford University, 382 Via Pueblo Mall, Stanford, CA 94305; \\
$^{10}$Department of Astronomy, University of Virginia, P. O. Box 400325, Charlottesville, VA 22904; \\
$^{11}$Department of Astronomy $\&$ Astrophysics, University of Toronto, 50 St. George Street, Toronto, ON M5S 3H4, Canada; \\
$^{12}$National Astronomical Observatory, Mitaka, Tokyo 181-8588, Japan; \\
$^{13}$Department of Physics and Astronomy, University College London, Gower Street, London, WC1E 6BT, UK; \\
$^{14}$Department of Physics and Astronomy, 6224 Agricultural Road, University of British Columbia, Vancouver, BC, V6T 1Z1 Canada; \\
$^{15}$Jeremiah Horrocks Institute of Maths, Physics and Astronomy, University of Central Lancashire, Preston, PR1 2HE, UK;
}

\maketitle

\footnotetext[1]{Nicholas Galitzki: E-mail: galitzki@sas.upenn.edu, Telephone: (215) 573-7558}


\begin{abstract}
The Balloon-borne Large Aperture Submillimeter Telescope for Polarimetry (BLASTPol) was a suborbital experiment designed to map magnetic fields in order to study their role in star formation processes. BLASTPol made detailed polarization maps of a number of molecular clouds during its successful flights from Antarctica in 2010 and 2012. We present the next-generation BLASTPol instrument (BLAST-TNG) that will build off the success of the previous experiment and continue its role as a unique instrument and a test bed for new technologies. With a 16-fold increase in mapping speed, BLAST-TNG will make larger and deeper maps. Major improvements include a 2.5 m carbon fiber mirror that is 40\% wider than the BLASTPol mirror and $\sim$3000 polarization sensitive detectors. BLAST-TNG will observe in three bands at 250, 350, and 500\micron. The telescope will serve as a pathfinder project for microwave kinetic inductance detector (MKID) technology, as applied to feedhorn coupled submillimeter detector arrays. The liquid helium cooled cryostat will have a 28-day hold time and will utilize a closed-cycle $^3$He refrigerator to cool the detector arrays to 270 mK. This will enable a detailed mapping of more targets with higher polarization resolution than any other submillimeter experiment to date. BLAST-TNG will also be the first balloon-borne telescope to offer shared risk observing time to the community. This paper outlines the motivation for the project and the instrumental design.
\end{abstract}


\section{Introduction} 
\begin{figure}[!bht]
  \centerline{
    \includegraphics[height=2.5in]{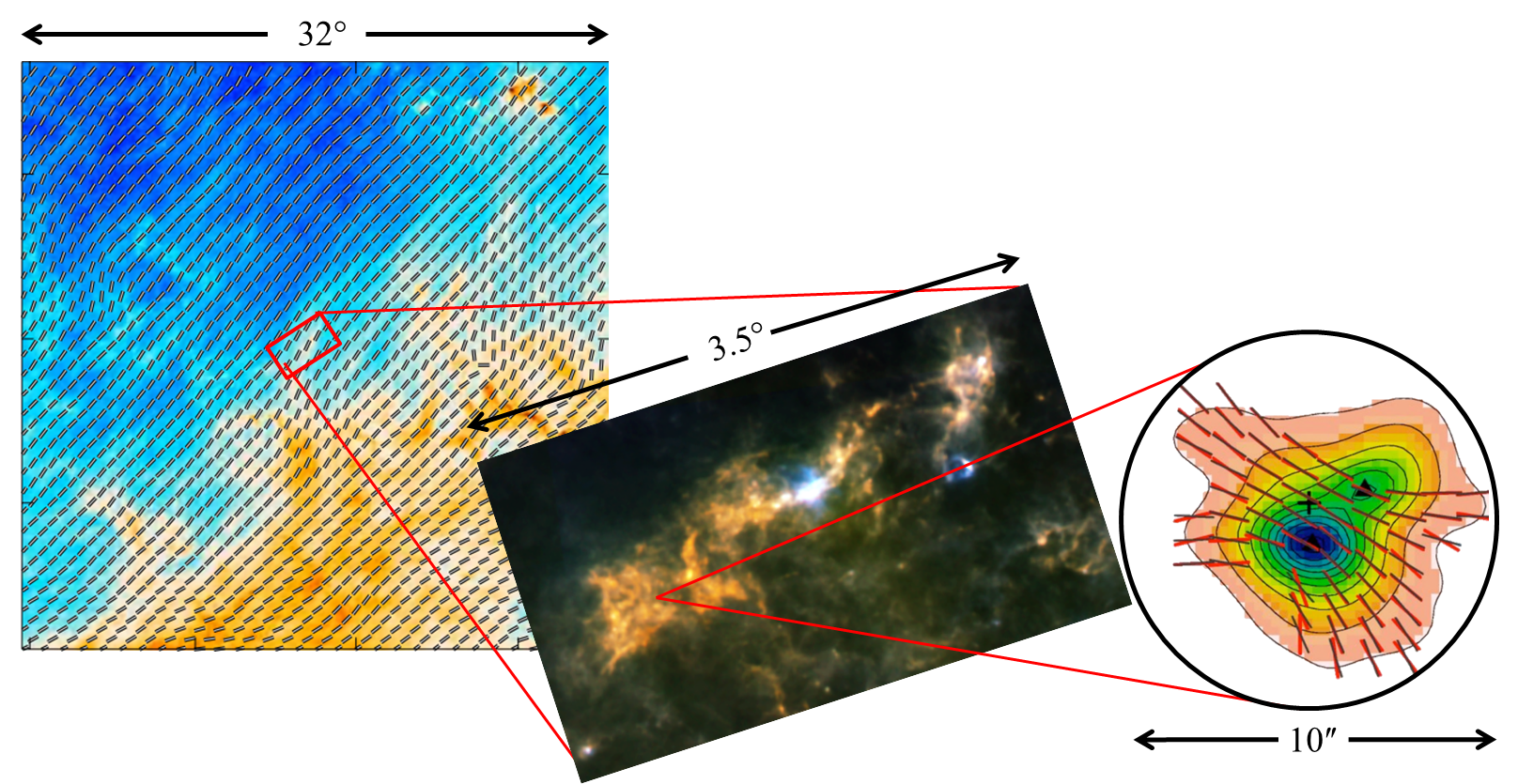}
  }
  \caption{\label{fig:scale_comp}BLAST-TNG provides an essential link between {\it Planck}'s all-sky polarization maps and ALMA's 0.01\arcsec\ resolution polarimetry. Left: A {\it Planck} polarization map of the Aquila Rift \cite{Planck2014} with 1$^\circ$ polarization resolution. Middle: The BLAST observation of the Vela molecular cloud \cite{Netterfield2009}. Right: A magnetic field map of a proto-binary in Persues from observations with the Submillimeter Array (SMA), a precursor to ALMA \cite{Girart2006}. By comparison, the BLAST-TNG beam at 250\micron\ is 22\arcsec, which is roughly the size of the ALMA 850\micron\ field of view and more than 200 times smaller in area than the {\it Planck} 850\micron\ beam.}
\end{figure}

BLAST-TNG fills an important gap in observational abilities of modern telescope facilities. BLAST-TNG will observe at wavelengths that are very difficult to observe from the ground, even from the best telescope sites in the world. It will make maps of high extinction molecular cloud regions where it is difficult to obtain IR polarimetry measurements from background stars. Additionally, the scale of the maps and resolution of the instrument provide an essential connection between the all sky, but low resolution (5\arcmin), polarimetric maps of {\it Planck} and the high resolution, but small area (22\arcsec), polarimetric maps of ALMA (see Figure \ref{fig:scale_comp}). The combination of data from all three telescopes will allow us to trace magnetic field orientation from the pre-stellar cores into the local molecular cloud and out into the surrounding Galactic environment. The design of the instrument will allow us to make deep maps of regions ranging from 0.25 to 20 square degrees and we will be able to observe many more star forming clouds than before.

BLAST-TNG will also target and characterize galactic dust polarization that is a source of foreground contamination for experiments that probe the Cosmic Microwave Background (CMB). Ongoing attempts to observe the polarized signal from the CMB \cite{BICEP22014,Polarbear2014,SPT2013,BICEP1_2014} rely on precise measurements of the polarized dust emission from our galaxy to extract the primordial signatures created during inflation. 

BLAST-TNG continues the legacy of BLASTPol \cite{Fissel2010} which has shown the effectiveness of submillimeter telescopes in making degree-scale maps of magnetic field pseudo-vectors from dust emission in large molecular clouds. The telescope will feature increased spatial resolution, a larger field of view (FOV), increased mapping speeds, and a longer flight time. BLAST-TNG will observe three 30\% fractional bandwidths centered on 250, 350 and 500\micron,  with 859, 407, and 201 pixels, having a beam FWHM of 25\arcsec, 35\arcsec, and 50\arcsec, respectively. The FOV of each array will be 22\arcmin. The detectors are feedhorn coupled MKIDs. Polarimetry is achieved with two orthogonal polarization sensitive MKIDs in each pixel along with a stepped Achromatic Half Wave Plate (AHWP) \cite{Moncelsi2014} that modulates the incoming light. The BLAST-TNG bands were chosen to be most sensitive to thermal emission from dust grains within molecular clouds. BLAST-TNG will fly from Antarctica in December of 2016 as part of NASA's Long Duration Balloon (LDB) program.

Section \ref{sec:scimo} describes the current state of scientific knowledge, Section \ref{sec:results} describes results obtained from previous BLAST experiments, and Section \ref{sec:instr} covers the details of the design of BLAST-TNG and the progress that has been made so far.

\section{Scientific Motivation}
\label{sec:scimo}

In recent years, understanding of the star formation process has advanced considerably. However, the mechanisms that reduce the rate of Galactic star formation are still poorly understood. Observational evidence points to star formation rates that are lower by up to a factor of 10 than those predicted by simple models based on free fall gravitational collapse. Current ideas to explain this discrepancy point to a support process that slows the formation and evolution of pre-stellar cores in molecular clouds. The two dominant theories that strive to explain this effect focus, respectively, on turbulent forces and magnetic fields \cite{McKee2007}. In the case of turbulence-controlled star formation, motion within the clouds dissipates over dense regions before they can reach a critical stage of collapse. Alternatively, trapped magnetic field lines may provide significant support against cloud collapse perpendicular to the field lines. In this case, collapse would be more rapid along field lines. In order to resolve the relative importance of these two competing mechanisms, we need information on velocity dispersions within clouds, obtainable from spectral line data, and magnetic field orientation and strength, observable with submillimeter polarimeters. 

Our efforts focus on determining magnetic field structure and correlation with molecular cloud features. Alternative methods for measuring local magnetic fields include Zeeman splitting, which is restricted to extremely bright regions, and optical extinction polarization observations, but these are limited to regions of low column density \cite{Crutcher2010,Falgarone2008}. For large extended molecular cloud regions with higher extinction, the most effective way to detect magnetically aligned grains is via far-IR and submillimeter polarimetry \cite{Hildebrand2000,WardThompson2000,Li2006,WardThompson2009}. Dust grains preferentially align with the local magnetic field and emit polarized light along their long axis which is orthogonal to the local magnetic field. Submillimeter polarimetry of molecular clouds measures the average polarization of light from a column of dust. By examining the dispersion of vector angles, combined with complementary observations of spectral line widths to determine a velocity dispersion in the cloud, the local magnetic field strength can be estimated using a method pioneered by Chandrasekhar and Fermi \cite{Chandrasekhar1953}. 

The precise mechanism that aligns the dust grains is not fully understood \cite{Lazarian2007}. Dust grain theories predict different sizes, shapes, and compositions that can have distinct effects on the polarization spectra. Multiband polarization measurements provide constraints for dust grain models.

Cosmic microwave background (CMB) observations have placed a large degree of importance on probing the era of inflation by observing the polarized signal to look for B-mode patterns as evidence of primordial gravity waves \cite{Smith2009}. In order to measure these fluctuations with a high degree of confidence, foreground Galactic dust contamination must be very well understood. Observing regions in the submillimeter that are being used by CMB polarization experiments will help to constrain foreground models, allowing for more detailed probes of inflation.

\section{Results from BLAST and BLASTPol}
\label{sec:results}

The Balloon-borne Large Aperture Submillimeter Telescope (BLAST) \cite{Fissel2010,Pascale2008,Marsden2008} flew from Kiruna, Sweden, in 2005 and from McMurdo, Antarctica, in 2006. It successfully served as a pathfinder mission for {\it Herschel}'s SPIRE instrument by flying and testing a similar detector and focal plane design. BLAST made a number of high profile observations, which included a high resolution map of the Vela C molecular cloud complex and confusion limited FIR observations of the GOODS South region \cite{Devlin2009,Marsden2009,Pascale2009,Patanchon2009,Viero2009,Wiebe2009,Netterfield2009}.

\begin{figure}[t]
  \centerline{
    \includegraphics[height=3.2in]{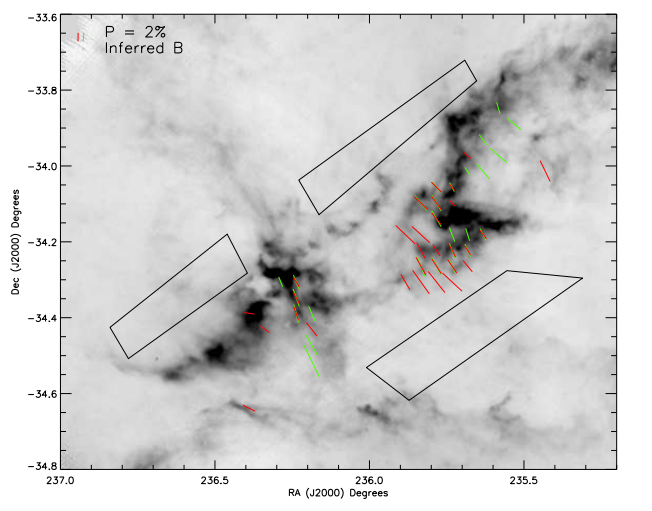}
  }
  \caption{\label{fig:lupus} Magnetic field pseudo-vectors obtained during the BLASTPol 2010 Antarctic flight \cite{Matthews2014}. The image is of the Lupus I star forming region with the intensity map provided by {\it Herschel} SPIRE 350\micron\  measurements. Boxed areas denote reference regions used in deriving the polarization pseudo-vectors which are then rotated by 90$^\circ$ to show inferred magnetic field pseudo-vectors. The length of the lines indicates the degree of polarization as dictated by the key in the upper left. Red and green pseudo-vectors show 500 and 350 micron measurements, respectively.}
\end{figure}

BLAST was then upgraded to become a polarimeter by inserting polarizing grids at the opening of the feedhorns and by adding an AHWP to the optical configuration. BLASTPol was able to make some of the first degree-scale polarization maps of nearby star forming regions. These maps cover multiple targets and contain thousands of pseudo-vectors [Fissel {\it et al.}, 2014, in preparation]. Its flights in 2010 \cite{Pascale2012} and 2012 [\citealt{Galitzki2014}, Angil\'e {\it et al.}, 2014, in preparation] proved the potential of this type of instrument to observe Galactic polarization. BLAST experiments also have a history of serving as test beds for balloon-borne telescope technology that have produced significant improvements in the field during its years in operation. An example of the type of observations obtained with these instruments is illustrated with a magnetic field map of the Lupus I cloud in Figure \ref{fig:lupus}.

\section{Instrument}
\label{sec:instr}

\begin{figure}[t]
  \centerline{
    \includegraphics[height=4in]{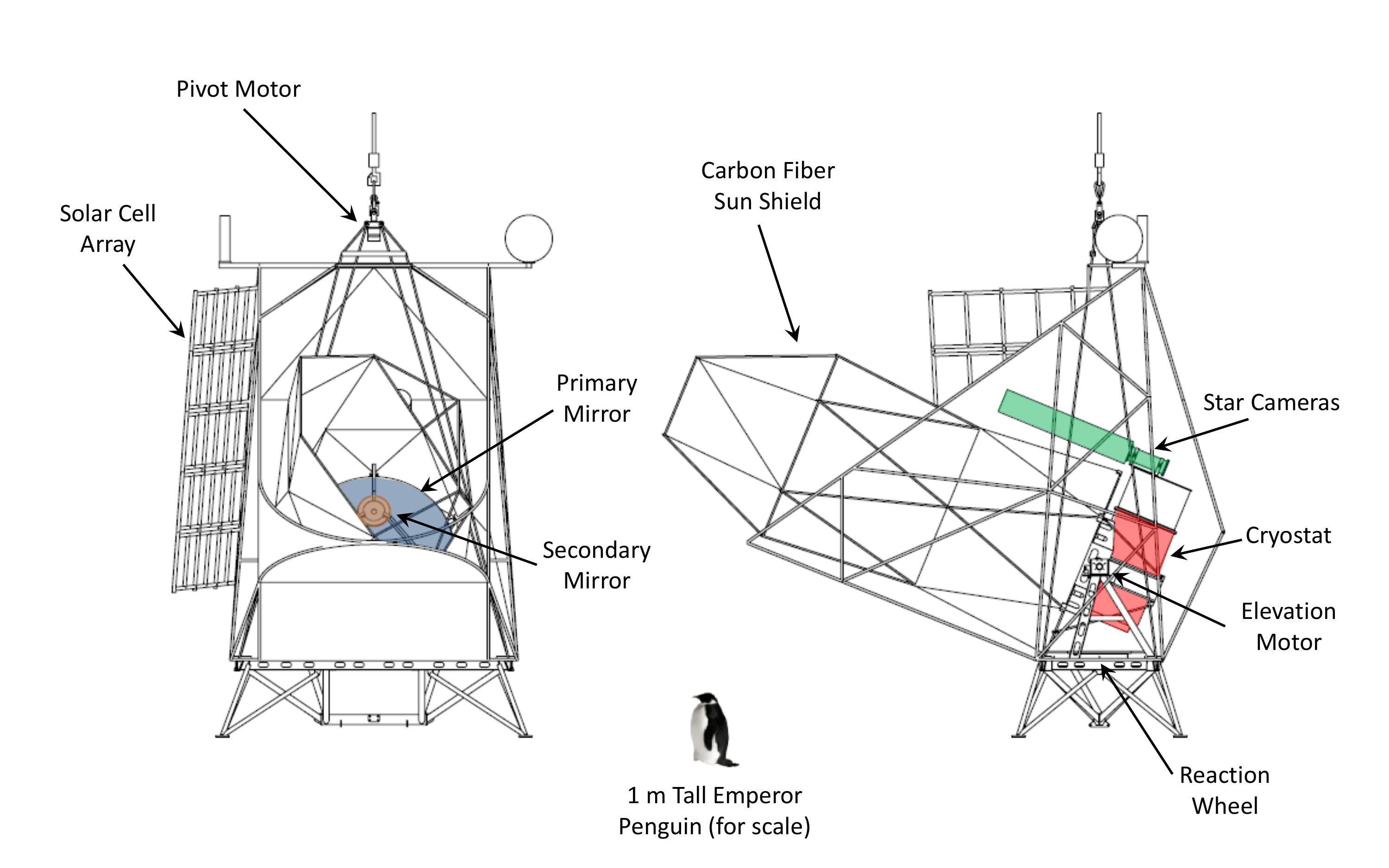}
  }
  \caption{\label{fig:model} The front and side views of the BLAST-TNG telescope in its flight configuration. The cryostat, mirror optics bench, and star cameras are attached to an inner frame that moves in elevation. An extensive carbon fiber sun shield also attaches to the inner frame to shield the optics at our closest pointing angle of 35$^\circ$ to the sun.}
\end{figure}

BLAST-TNG will continue the legacy of BLAST and BLASTPol with the construction of an entirely new instrument. The design incorporates many successful elements from previous ballooning experiments. It is based around a 2.5 meter primary mirror which provides diffraction-limited observations in three 30\% fractional bands at 250, 350, and 500 \micron. The FOV has been expanded to 22\arcmin\ in diameter. This has led to a proportional increase in the size of the optics necessitating the construction of a new cryostat. The cryostat has been designed to have a longer hold time of 28 days, versus the 13 day hold time of BLASTPol. The previous cooling system used both helium and nitrogen, however, for BLAST-TNG we have switched to an entirely liquid helium system that utilizes two vapor-cooled shields to provide additional thermal isolation. The instrument will also serve as a pathfinder instrument for MKIDs \cite{Day2003}, which have never been flown before. The development of MKID arrays for astronomy is an extremely active area of detector research, and flight testing them will be a significant milestone. The design of the instrument is shown in Figure \ref{fig:model}.

\blfootnote{{\it Herschel} is an ESA space observatory with science instruments provided by European-led Principal Investigator consortia }
\blfootnote{and with important participation from NASA.}

\subsection{Pointing Systems and Electronics}
\label{ssec:point}

A number of the BLAST-TNG pointing and electronics systems are inherited from or based off components successfully flown previously with BLAST and BLASTPol. We will be using many of the same sensors to determine telescope attitude, such as the star cameras, gyroscopes, and sun sensors \cite{Gandilo2014}. Pointing of the telescope will use the same strategy of a precision pointed inner frame that moves in elevation, supported by an external gondola that scans in azimuth. However, the computer and electronics systems will undergo a major overhaul to reduce reliance on legacy components that are either outdated or difficult to maintain. 

\subsubsection{Pointing System}

The pointing in azimuth is controlled by a reaction wheel and a pivot motor. The reaction wheel is a 1.5 meter diameter wheel with high angular moment of inertia that is kept in motion by a brushless direct drive motor. Adjusting the speed of the wheel results in a transfer of angular momentum from the wheel to the gondola and allows us to scan in azimuth at typical rates of 0.05 to 0.2$^\circ/$s while observing, and slew at speeds of several degrees per second while moving between targets. The pivot motor, which is attached between the payload and the flight train, can assist the reaction wheel, if its speed begins to saturate, by transferring angular momentum to the balloon. The pivot motor can also independently point the telescope in azimuth in case of reaction wheel failures, as occurred during the 2010 flight due to a bearing malfunction in the reaction wheel motor.

Pointing in elevation was previously accomplished by a direct drive brushless servo motor that was connected to the axis of an inner frame, on which the telescope and cryostat were mounted. This worked in conjunction with a balance system that periodically pumped fluid from the bottom to the top of the inner frame to counteract the imbalance caused by cryogens boiling off. With the much larger cryogenics volume of BLAST-TNG and the overall more massive structure, we have decided to couple the direct drive motor to a Harmonic Drive\footnote{Harmonic Drive LLC: 247 Lynnfield Street, Peabody, MA 01960} gear-head with a ratio of 80:1. This approach eliminates the need for a balance system, which is preferred due to the complications of operating a fluid system in near vacuum conditions.

The primary absolute pointing sensors are two bore-sight star cameras that are able to determine the right ascension (RA) and declination (Dec) of the instrument to $<$ 2\arcmin\ during the flight and $<$ 5\arcsec\ after post-flight reconstruction of the pointing solution \cite{Rex2006}. Each star camera contains a high-resolution (1 megapixel) integrating CCD camera with a 200 mm f/2 lens, to image a 2$^\circ$ x 2.5$^\circ$ FOV. A stepper motor controls the focus, to compensate for thermal variations, while another stepper motor controls the aperture, to allow for different exposure times, depending on the scan strategy. Each camera is connected to a PC-104 computer that uses either a `lost-in-space' algorithm, that searches the entire sky for the position, or an algorithm developed for BLAST, that incorporates information from the coarse sensors and previous pointing solutions to find the new position. The latter process is needed to reduce the time it takes for the cameras to produce a pointing solution. Every time the star cameras capture an image, they send the pointing information to the flight computers along with image data that can be used in the post flight pointing reconstruction. There are two star cameras to provide more frequent pointing information as well as redundancy should one of them fail, as occurred in the 2012 flight. 

Two sets of three DSP-3000 fiber optic gyroscopes\footnote{KVH Industries, Inc.: 50 Enterprise Center, Middletown, RI 02842} provide fast relative solutions, but have a drift of 1\arcsec/s. They are primarily used to interpolate the relative pointing between star camera solutions. There are a pair of redundant gyroscopes aligned with each major axis, with relative offsets determined prior to flight. 

BLAST-TNG will fly with additional coarse pointing sensors (see Table \ref{tab:pointing}) that serve as complimentary sensors to the star cameras in case they malfunction or experience difficulty obtaining solutions. Pin hole Sun sensors use the Sun's location in the sky to determine pointing in azimuth and were developed and tested in previous flights \cite{Korotkov2013}. A magnetometer determines azimuth pointing information by measuring the orientation of Earth's magnetic field. There are also two inclinometers that determine the tilt of the inner and outer frames which can be used to measure the pointing elevation. We will also be including a precision optical elevation encoder mounted between the elevation motor gear-head and the inner frame. We plan to use a RESOLUTE absolute rotary encoder on RESA\footnote{Renishaw: 5277 Trillium Blvd, Hoffman Estates, IL 60192} rings which will have over 10 times the resolution of the previous encoder used.

Our philosophy with the pointing system is to have a degree of redundancy for most all components in case of a critical failure. Our combination of fine and coarse sensors allows us to achieve our in flight pointing requirements. During post flight pointing reconstruction, we have achieved accuracy $<$ 5\arcsec\ RMS \cite{Gandilo2014}, which is more than adequate for our observations.
\begin{table}[t]

\newcommand\T{\rule{0pt}{3ex}}       
\newcommand\B{\rule[-2ex]{0pt}{0pt}} 
\vspace{3 mm}
\begin{center}
\caption{Summary of Pointing Sensor Parameters \cite{Gandilo2014} \label{tab:pointing}}
\vspace{3 mm}
\renewcommand{\arraystretch}{1.2}%
\begin{tabular}{lcc}
\hline
\hline
\T Sensor & Sample Rate (Hz) & Accuracy ($^\circ$) \B \\
\hline
\T GPS & 10 & 0.1 \\   
Sun Sensor & 20 & 0.1 \\
Magnetometer & 100 & 5 \\
Clinometer & 100 & 0.1 \\
Star Camera & 0.5 & $<$0.001 \\
Elevation Encoder & 100 & $<$0.001 \B \\                                                    
\hline
\end{tabular}
\end{center}
\end{table}

\subsubsection{Electronics}

\begin{figure}[t]
  \centerline{
    \includegraphics[height=3.2in]{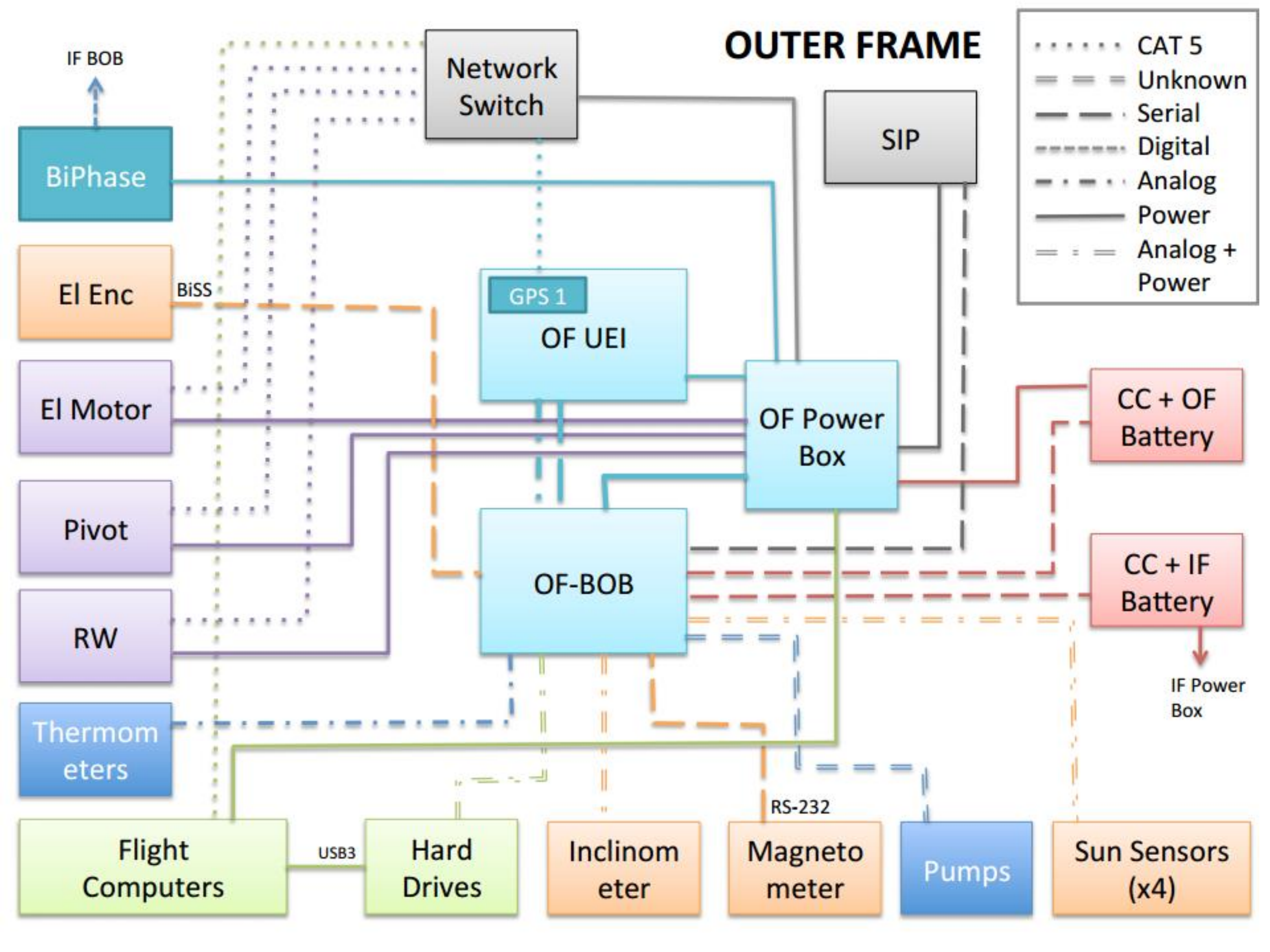}
  }
  \caption{\label{fig:electronics} A schematic of the electronics system on the outer frame (OF) that shows all major components and how they connect. The hard drives are kept in a separate pressurized vessel that is connected to the computers. The pivot, reaction wheel, and elevation motors all have separate motor controllers that receive commands over ethernet. The OF electronics are connected to a similar system on the inner frame (IF) that runs the cryostat motors, secondary mirror actuators, and electronics associated with the cryostat and detectors. The SIP and BiPhase are components provided by NASA for controlling their equipment and the communications systems.}
\end{figure}

Ballooning systems require interfacing numerous pointing sensors and motors to provide effective attitude control. The previous instrument relied on a flight computer running a master program that handled I/O through a PCI BLASTBus card \cite{Benton2014}. This system was developed by the original BLAST collaboration and has successfully been used in a number of experiments \cite{Rahlin2014,Oxley2004,Swetz2011,Ogburn2010,Staniszewski2012}. However, the BLASTBus has become impractical to use, given that much of the expertise and spare parts are no longer available. The BLAST-TNG instrument will utilize a new commercial system for applications in balloon-borne missions. We will be using two United Electronics Industries (UEI)\footnote{United Electronic Industries: 27 Renmar Avenue, Walpole, MA 02081} DNA-PPC8 cubes to handle sensor I/O, thermometry, motor control, time synchronization, and other functions. The UEI cubes are very flexible, with a variety of interchangeable cards that can be tailored to our specific needs. We will be using a combination of ADC, DAC, DIO, RS-232/485 8-port serial, and timing control/GPS cards. Each cube has an embedded PPC CPU running Linux 3.2 with real-time extensions that will run a custom control system written in C.  Synchronization between systems is handled by the dedicated UEI sync interface port. 

The power system is divided into two components, one which provides power to the flight computers and outer frame electronics, and another which provides power to the detector readout system and inner frame electronics. Each power network will be charged by six SunCat solar panels\footnote{SunCat Solar, LLC: HC 1 Box 594, Elgin, AZ 85611}. At 28 V, the solar panels can provide over 1100 Watts of normal incidence power and over 500 Watts at an incidence angle of 60 degrees. The solar array output for each power network is routed through a Morningstar TriStar 60 amp MPPT solar charge controller\footnote{Morningstar Corporation: 8 Pheasant Run, Newtown, PA 18940} (CC), which maintains the proper voltage to charge the batteries and can be monitored and controlled over a serial line. The charge controller couples to two 14 volt lead-acid batteries connected in series.  The lines from the batteries feed into a power box, which converts the battery voltage into the various voltages required by the instrument while also providing switching. 

The pivot, reaction wheel, and elevation motors that control the telescope pointing will utilize Copley motor controllers\footnote{Copley Controls: 20 Dan Road, Canton, MA 02021} and are commanded via EtherCAT, a network-based CANBus protocol. The signal cables will be routed through two breakout boxes (BOB) that redistribute cables from the UEI cube cards to their respective destinations. 

One UEI cube, BOB, and power box are mounted to the inner frame along with the detector readout and cryostat electronics, while another set is attached to the outer frame, along with the flight computers and hard drives. The outer frame configuration is shown in Figure \ref{fig:electronics}. The UEI cubes greatly simplify the design and eliminate the need for the BLASTBus system. All components are currently being programmed and tested for preliminary integration with pointing sensors in summer 2014.

\subsection{Cryogenics}
\label{ssec:cryo}

\begin{figure}[t]
  \centerline{
    \includegraphics[height=3.5in]{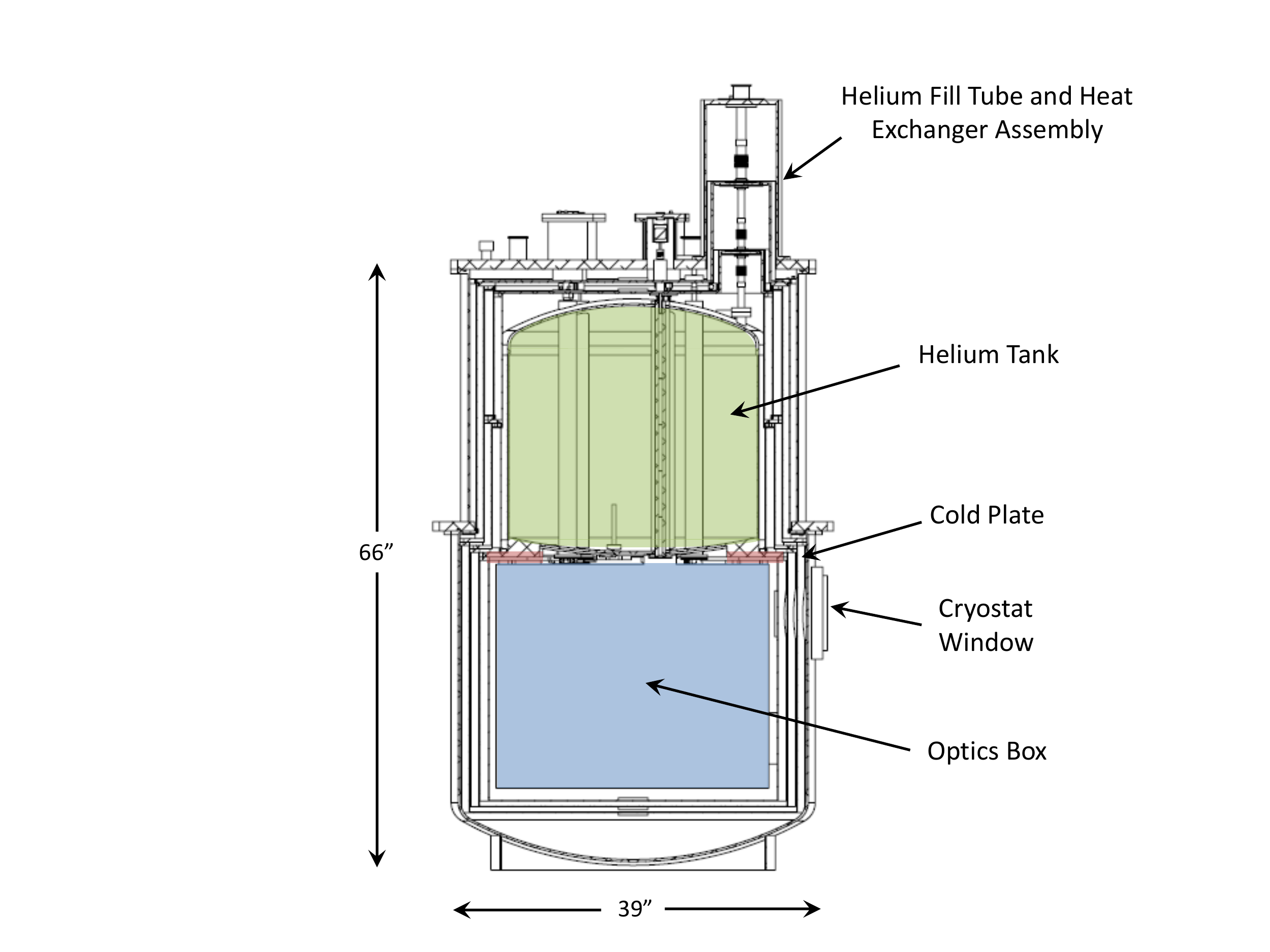}
  }
  \caption{\label{fig:cryostat} Cross section view of the BLAST-TNG cryostat. The cryostat is cooled by liquid helium with two helium vapor heat exchangers that cool two shields to provide thermal isolation of the cold optics. The inner vapor cooled shield is kept at 66\,K and the outer one at 190\,K. The cryostat has a predicted hold time of 28 days.}
\end{figure}

The detector arrays nominally operate at 270 mK, which requires a robust cooling system to reach the required temperature and to keep the temperature steady during normal flight operations. In order to simplify the design and cost of the cryostat, we are not using liquid nitrogen as an intermediate cooling stage, but are instead using a system of two vapor-cooled shields. These shields are cooled by a heat exchanger that extracts the enthalpy of the gas, as it warms from 4\,K to 300\,K. Preliminary tests on a prototype heat exchanger suggest efficiencies between 80\% and 90\%. Our cryostat thermal models use a baseline efficiency of 80\% for the heat exchangers to calculate a hold time in excess of 28 days. The cryostat is currently under construction at Precision Cryogenics \footnote{Precision Cryogenics Systems, Inc.: 7804 Rockville Road, Indianapolis, IN 46214}.


The helium tank required an innovative design to accommodate the 250\,L of liquid helium while also providing a stable mounting platform. The cold optics box is bolted directly to the helium tank to minimize the thermal path between the optics and the helium bath. The cold optics box is also mounted to the rim of the helium tank to minimize the effect of the deflection of the cold plate on the precision optics. The structure of the helium tank is further reinforced by six feed-through tubes that allow for easy insertion of cabling, piping, and motor axles from the top of the cryostat to the cold plate (Figure \ref{fig:cryostat}).

For the low temperature cryogenics system, we will reuse the helium refrigerator from BLASTPol with only minor modifications. The system consists of a pumped $^4$He pot and a $^3$He refrigerator. The $^4$He pot functions by pulling a vacuum on a small reservoir, $\sim$200\,mL, that is periodically filled, from the main tank, through a small capillary tube, with flow controlled by a motorized valve. The $^4$He pot cools to around 1.2\,K, which serves as an intermediate stage for the thermal isolation of the detectors and as part of the cycling process of the $^3$He refrigerator. The closed cycle $^3$He refrigerator has a small $^3$He reservoir that utilizes evaporative cooling to reach temperatures around 270\,mK. The vapor is then adsorbed by a sphere of charcoal to maintain low pressure conditions in the refrigerator. To cycle the refrigerator, the charcoal is heated to release the $^3$He as vapor, which condenses on a section cooled by the $^4$He pot and refills the reservoir. Each cycle lasts 90 minutes and we predict we will need to cycle every $\sim$40 hours under maximum loading of the 270\,mK stage.

\subsection{Optics}
\label{ssec:opt}
\begin{figure}[t]
  \centerline{
    \includegraphics[height=3in]{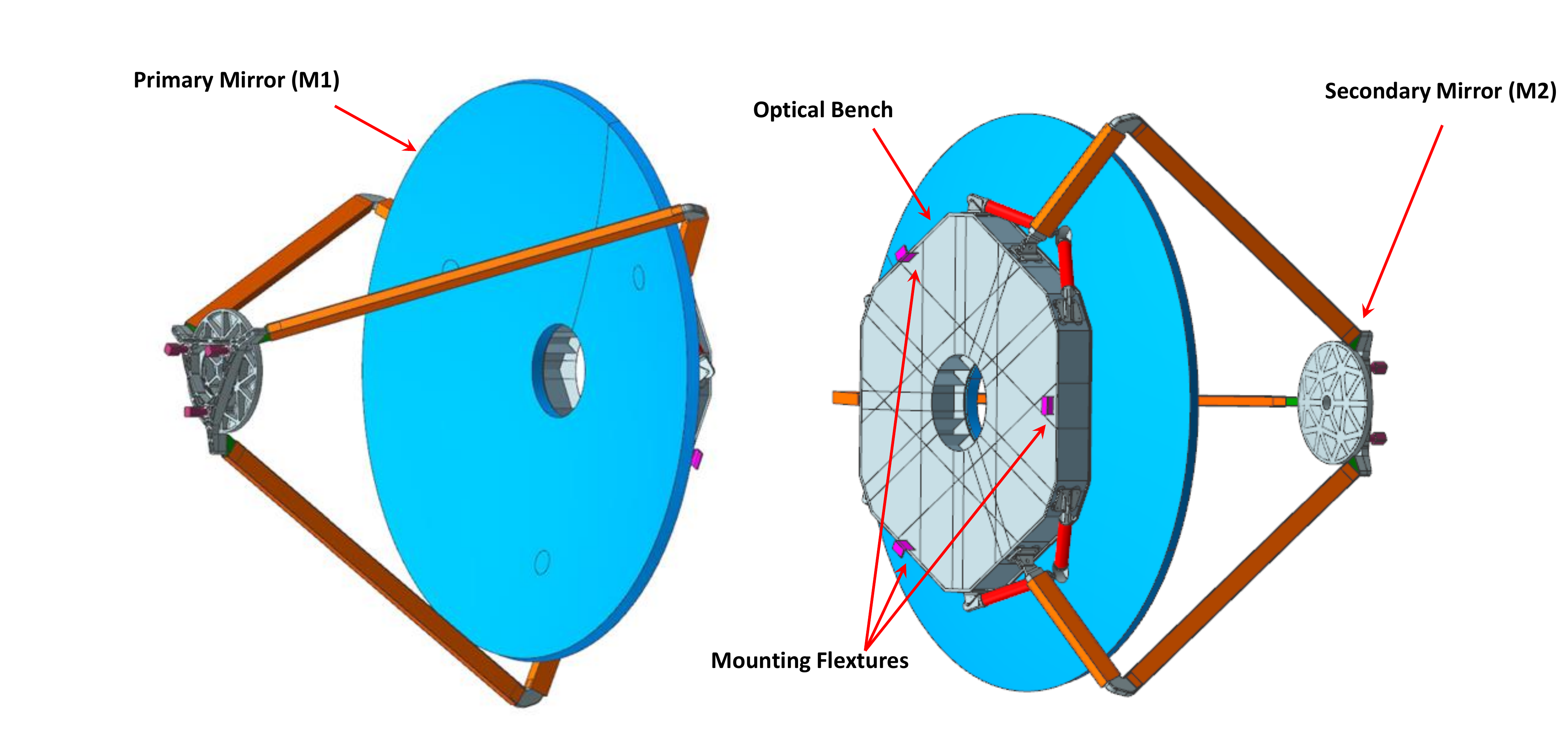}
  }
  \caption{\label{fig:mirror} A design concept from Vanguard Space Technologies for the primary and secondary mirror structure. The primary is a 2.5 meter diameter carbon fiber parabolic mirror while the secondary is a 52 cm diameter aluminum hyperbolic mirror that will be actuated with respect to the primary to allow for in-flight focusing. Both mirrors attach to a backing optical bench that is made of carbon fiber and will mount to the inner frame of the gondola.}
\end{figure}

The primary and secondary mirrors of the BLASTpol instrument were arranged in a Ritchey-Chr{\'e}itien design with a 1.8 m diameter aluminum primary (M1) and a 40 cm diameter aluminum secondary (M2). These were originally connected by four carbon fiber struts, but it was found before the 2012 campaign that the support structure distorted M1 where they attached to the mirror's edge. A new system was developed with three equally spaced struts that attached to the inner frame. This layout greatly reduced the stress on M1. BLAST-TNG will have a Cassegrain configuration with a 2.5 m carbon fiber reinforced polymer (CFRP) M1 and a 52 cm diameter aluminum M2. M1 and the three CFRP struts that support M2 are attached to a rigid CFRP optical bench that serves as a backing structure and interface to the gondola inner frame. The bench and both mirrors are being developed and built by Vanguard Space Technologies\footnote{Vanguard Space Technologies: 9431 Dowdy Drive, San Diego, CA 92126} through a NASA Small Business Innovation Research (SBIR) grant. The primary mirror is expected to have better than $\sim$10\micron\ RMS surface error under operating conditions. A conceptual image of the M1 and M2 structure is shown in Figure \ref{fig:mirror}. The mirror is scheduled to be completed by the end of 2015.

\begin{figure}[t]
  \centerline{
    \includegraphics[height=2.5in]{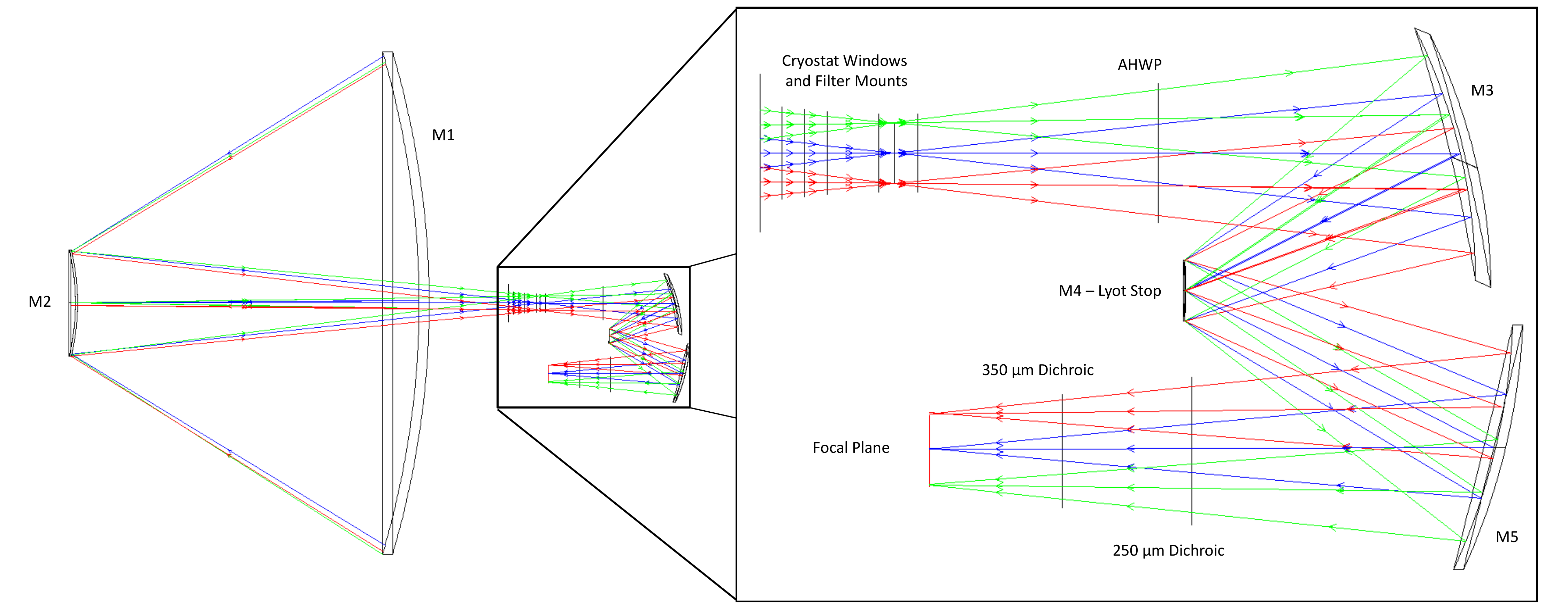}
  }
  \caption{\label{fig:optics} Side view of the optical configuration of BLAST-TNG with a detailed view of the cold optics. The telescope is an on axis Cassegrain that feeds into a modified Offner relay. M3, M4, and M5 are spherical mirrors with M4 acting as the Lyot stop for the telescope with a blackened hole that shadows the secondary mirror. Within the hole at the center of M4 is a calibrator lamp that provides an absolute calibration during flight operations to monitor responsivity drifts in the detectors. There are two dichroics that split the beam to the 250 and 350\micron\ arrays. The 250\micron\ dichroic is tilted at 22.5$^\circ$ to the optical axis while the 350\micron\ dichroic is tilted at 30$^\circ$ to the optical axis. Only one of the three focal planes is shown. The AHWP is inserted between the Cassegrain focus and M3.}
\end{figure}

The cold optics for BLAST-TNG will be a larger version of what was used on BLASTPol to accommodate the wider FOV. The cold optics will be at 4\,K and are in a modified Offner relay configuration (see Figure \ref{fig:optics}). Three spherical mirrors, M3, M4, and M5, refocus the beam from the Cassegrain output to the detectors with an adjusted f/\# such that the beam is f/5 at the focal plane. The M4 mirror is the Lyot stop, which under-fills the primary mirror to 2.4 meters and has a hole in its center to shadow the secondary mirror. A calibrator lamp is mounted in the hole in the middle of M4 such that it evenly illuminates the entire FOV when pulsed. The lamp was developed for the {\it Herschel} SPIRE instrument and was specifically designed for the calibration of submillimeter detectors \cite{Pisano2005}. The two main improvements to the BLASTPol design are removal of flat, beam folding mirrors, which allows easier access to the arrays, and mounting all components to a single optical bench for ease of assembly and alignment. The optical parameters are listed in Table \ref{tab:optica}.

\begin{figure}[t]
  \centerline{
    \includegraphics[height=2.5in]{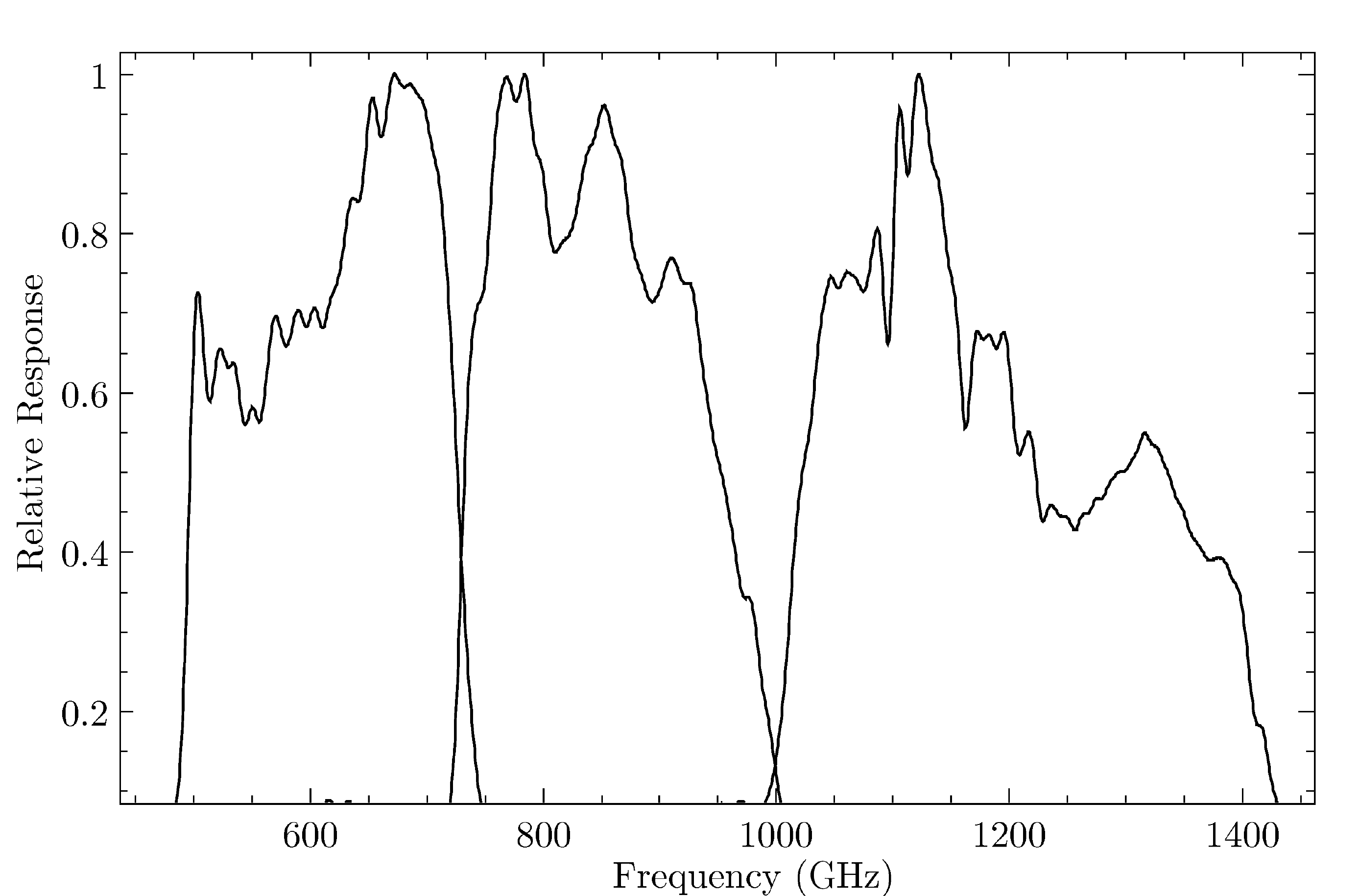}
  }
  \caption{Spectral response of all three BLASTPol bands normalized to peak transmission. Measurements were made with a Fourier Transform Spectrometer using a nitrogen source. Some atomspheric lines are present as tests were done at sea level with a small air gap between the cryostat and the FTS \cite{Wiebe2008}. 
    \label{fig:bandpass}}
\end{figure}

\begin{table}[t]

\newcommand\T{\rule{0pt}{3ex}}       
\newcommand\B{\rule[-2ex]{0pt}{0pt}} 
\vspace{3 mm}
\begin{center}
\caption{Summary of BLAST-TNG Optics Characteristics\label{tab:optica}}
\vspace{3 mm}
\renewcommand{\arraystretch}{1.2}%
\begin{tabular}{lccccc}
\hline
\hline
\T Geometrical Charac. & M1 & M2 & M3 & M4 & M5 \B \\
\hline
\T Nominal Shape & Paraboloid & Hyperboloid & Sphere & Sphere & Sphere\\                                                       
Conic Constant & $-$1.0 & $-$2.182 & 0.000 & 0.000 & 0.000\\
Radius of Curvature & 4.161$\,$m & 1.067$\,$m & 655.6$\,$mm & 376.5$\,$mm & 749.4$\,$mm\\
Aperture & $\varnothing$2.5$\,$m & $\varnothing$0.516$\,$m & $\varnothing$28$\,$cm & $\varnothing$7$\,$cm & $\varnothing$28$\,$cm \B \\
\hline
\end{tabular}
\end{center}
\end{table}
Prior to the light entering the cold optics, it must pass through a series of IR blocking filters and low pass edge filters \cite{Ade2006}, developed at Cardiff University, that are attached to the windows in the cryostat shells. These serve to reduce thermal loading on the cryostat and the detectors. The beam is split by two low pass edge dichroic filters placed after M5 to allow simultaneous observations in all three bands. The observed band is further constrained by filters mounted to the front of the arrays and by the feedhorn design. A measurement of the BLASTPol bands is shown in Figure \ref{fig:bandpass} \cite{Wiebe2008}.

\subsection{Detectors}
\label{ssec:mkid}

BLAST-TNG will serve as a pathfinder instrument with the first use of MKIDs on a balloon-borne platform. Each feedhorn-coupled pixel will have two orthogonally oriented detectors to simultaneously sample both the Q and U Stokes parameters that define linearly polarized light. The total number of pixels will be $\sim$1500 with $\sim$3000 MKID detectors, which is about 10 times the number of bolometric detectors flown with BLASTPol. 

MKID detectors have been identified as a promising new technology for astronomy, with potential applications from the submillimeter to the X-ray. The superconducting, titanium-nitride, MKIDs that are being designed for BLAST-TNG use a single loop inductor and interdigitated capacitor to form an LC circuit with a tuned resonant frequency \cite{Day2003}. Photons incident on the inductor, with energies greater than the gap energy, break Cooper pairs which causes a measurable change in the impedance of the inductor. The inductor loop is long and narrow to make it sensitive to a single direction of linearly polarized light. Dual-polarization sensitivity is achieved by placing two orthogonal detectors in a single feedhorn coupled pixel. Fabrication of the MKIDs is quite simple, with the primary circuit done in a single layer on a silicon wafer. Crossovers for the intersecting inductor loops are then added, made with silicon-oxide insulators and superconducting niobium bridges. Alternative detector technologies, such as transition edge sensor (TES) arrays, have considerably more complicated fabrication processes. Additionally, thousands of MKID signals can be multiplexed on a single feed-line, with the number of the detectors currently limited by the warm readout electronics technology. The signal from the MKIDs requires a single low power ($<$ 5 mW), wide-band, silicon-germanium amplifier, that operates at 4\,K. These factors make the production of large scale, dual-polarization, MKID detector arrays achievable on the time-scale of the BLAST-TNG project.
\begin{figure}[t]
  \centerline{
    \includegraphics[height=2in]{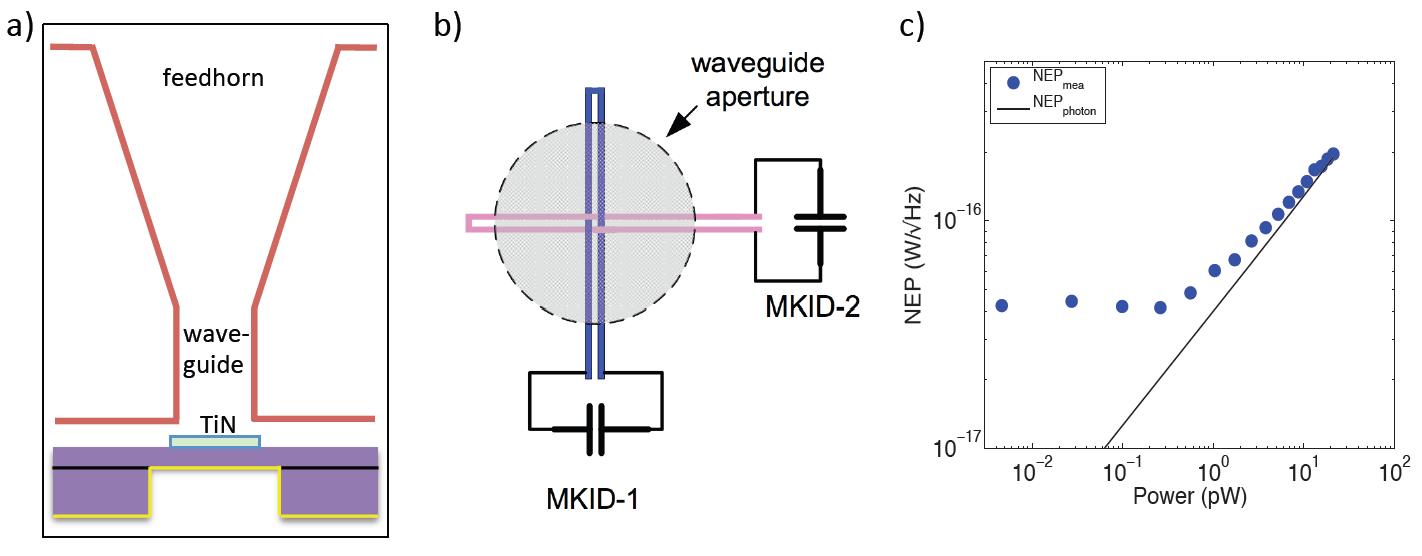}
  }
  \caption{\label{fig:mkid} {\bf (a)} A cross section of a single pixel with the feedhorn coupled MKID detector. {\bf (b)} A schematic of the detector layout with two single loop inductors orthogonally aligned at the end of the waveguide. Each inductor is part of an LC circuit with a set frequency that couples to a readout line. {\bf (c)} The measured noise equivalent power (NEP) of a BLAST-TNG prototype detector as a function of radiative load at a band centered on 250 micron \cite{Hubmayr2014}.  The data (blue points) are limited by photon noise (black line) at thermal loads above $\sim$1 pW, which includes our expected in-flight loading condition (7 to 17 pW).}
\end{figure}

\begin{figure}[t]
  \centerline{
    \includegraphics[height=2.5in]{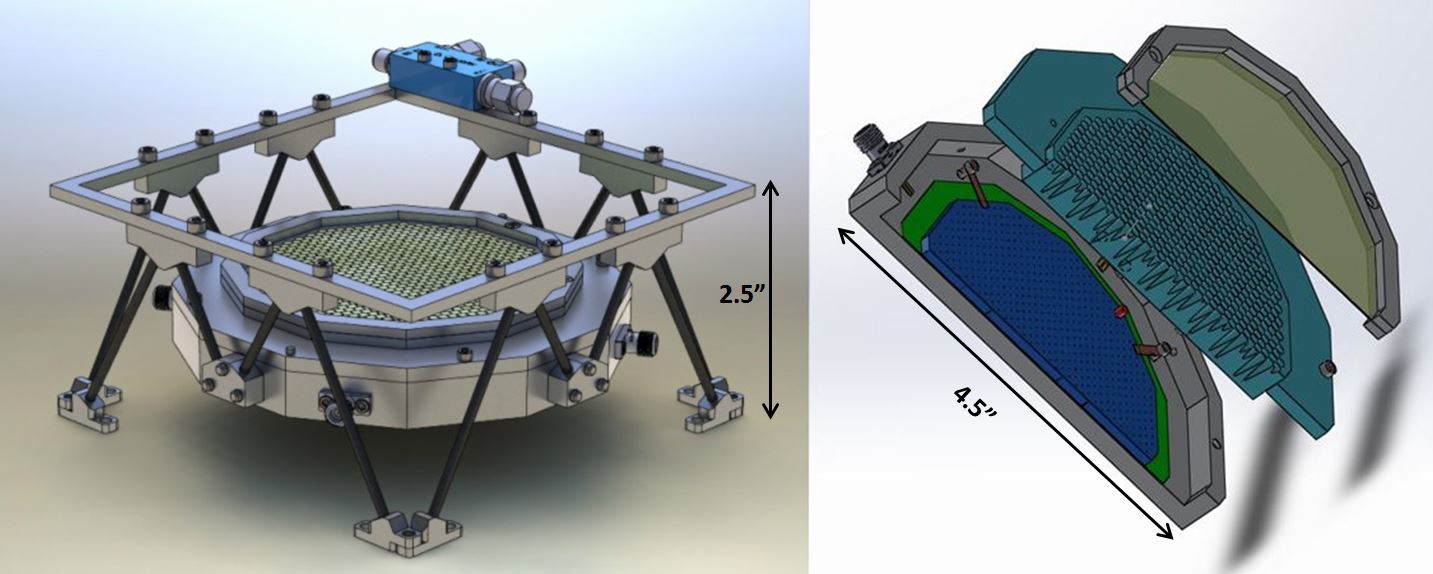}
  }
  \caption{\label{fig:fpa} Left: The complete focal plane array (FPA) support structure. The feet mount to fixtures that are attached to the optics bench. Carbon fiber rods provide thermal isolation by offsetting the FPA from the 4K optics bench via an intermediate 1K stage. Right: A cross section of the structure which includes from top to bottom: bandpass filter, feedhorn array, waveguide wafer (blue), detector wafer (green),  and FPA housing.}
\end{figure}

A precision machined aluminum feedhorn array is placed in front of the detector wafer to couple the light to the absorbing elements. The BLASTPol feed array used a conical feedhorn design, similar to the {\it Herschel} SPIRE feedhorns \cite{Rownd2003}. However, these were not optimized for polarimetry and have a divergence in the E and H fields that results in an asymmetry between the polarization directions. BLAST-TNG will use a modified Potter horn design \cite{Potter1963} with three steps, which excites additional modes in the EM field and reduces asymmetries in the polarized light, while maintaining the 30\% fractional bandwidth required \cite{Tan2012}. They also have the advantage of being much easier to fabricate than equivalent corrugated feedhorn designs.

A seven element feedhorn coupled detector array has been constructed and is being tested with a variable temperature black-body load at the National Institute of Standards and Technology (NIST). The tests aim to characterize the noise performance of different detector models in order to find the design that best meets the needs of the instrument. Recent results demonstrate that the current design for our detectors are photon noise limited over the range of loading levels expected during flight \cite{Hubmayr2014} (see Figure \ref{fig:mkid}). Current efforts are focused on demonstrating detector noise below 0.01-0.1 Hz, which is a rate determined by our scan speed across targets. 80\% co-polar and $<$1\% cross-polar absorption in a pixel has been predicted by running a high frequency structural simulator (HFSS) on a model of orthogonal absorbers with silicon-oxide insulated crossovers and a quarter wavelength deep metalized backshort. 

The focal plane array (FPA) is made of an eight-sided polygon housing, the detector wafer, the waveguide wafer, and the feedhorn block (see Figure \ref{fig:fpa}). The housing has mounts for the coaxial SMA connectors for the multiplexed feed-line, the thermal standoffs, and for the copper heat strap from the $^3$He refrigerator.  The detector and waveguide wafers are mounted using a combination of fixed pins and pins-in-slots to constrain the wafers in the plane of the array without causing stress from thermal contraction between different materials. The wafers are held against the housing by beryllium-copper spring tabs. There is extra space at the edge of the detector wafer to give room for wire bonds to the feed-line that also serve to thermally sink the detector array. The feedhorn array attaches to the rim of the housing and is offset from the waveguide wafer by a small gap ($\sim$20\micron). The feedhorn block also has a mounting rim for filters and alignment fixtures. 

The whole FPA assembly is thermally isolated by a carbon fiber structure that connects the FPA to a fixture at 1\,K, which then connects to the 4\,K optics bench. The input SMA cables connect to a directional coupler which serves to heat-sink the cables before they go into the arrays. The output SMA cables go directly to the cryogenic amplifier. Thermal modeling predicts loads from the structure, cables, and radiation to be $<20\mu$W and $<150\mu$W on the 270 mK and 1 K stages, respectively. Finite element analysis of the mechanical structure predicts deflections of $<$5\micron\ across the array under typical flight stresses. The detector and FPA designs are in an advanced stage of development and the final flight detector arrays will be completed by the end of 2015.

The FPAs will be shielded by an Amumetal 4K (A4K) box\footnote{The MuShield Company: 9 Ricker Ave., Londonderry, NH 03053} that encompasses the cold optics, which will significantly reduce the effect of local magnetic field fluctations on the MKID arrays. The box surrounds all of the cold optics to block stray light within the 4K shield and to reduce the quantity and size of the holes in the magnetic shield, with the significant exceptions being the window and feedthroughs for cabling.

\subsection{Detector Readout}

MKIDs have the advantage of a simple cryogenic layout, with a single-layer detector wafer and one coax line able to read out hundreds of detectors. However, this shifts a large amount of complexity to the warm readout electronics. BLAST-TNG plans to use Reconfigurable Open Architecture Computing Hardware -2 (ROACH-2) \cite{Werthimer2011} boards developed by the Collaboration for Astronomy Signal Processing and Electronics Research (CASPER), coupled with the MUSIC-developed ADC/DAC card and intermediate-frequency (IF) board \cite{Golwala2012} , to generate a frequency comb consisting of all the resonant frequencies of the detectors sampled by that system. The comb is then sent through coax to the feed-line that runs across the detector wafer. The modulated output frequency comb is boosted by a SiGe amplifier, provided by P. Mauskopf, which operates at 4\;K. After leaving the cryostat, the comb is digitized, analyzed, and compared to the input comb to identify shifts in phase of individual detectors due to impedance changes in the inductors. The relative shifts are time-stamped and then sent to the computer to be merged with pointing data, which is then recorded on the hard drives.

The central component in the readout chain is the ROACH-2 board. The ROACH-2 consists of a Xilinx Virtex 6 Field Programmable Gate Array (FPGA) coupled to two ZDOK connectors, a PowerPC CPU connected to 1Gb Ethernet, 72-bit DDR3 RAM for slow memory access, and  4 x 36 bit wide 288 Mb QDR II+ SRAMs for fast memory access. The ZDOKs are connected to the ADC/DAC card and are used to send out and read in the frequency combs. The FPGA is the workhorse of the ROACH-2 board and performs all the necessary digital signal processing (DSP) before sending out the packetized phase streams to the CPU to be sent out over the 1Gb Ethernet to the flight computer.

Since our MKID resonance  frequencies of 700\;MHz-1.25\;GHz are above the operational ADC/DAC frequency band of $\sim$0-500\;MHz, we require a local oscillator (LO) to convert the output tones from the baseband of the ADC/DACs to the resonance frequencies of the MKIDs and vice-versa. We perform these conversions via two IQ-mixers. These mixers also allow us to stitch together both 500 MSPS ADCs into a single 500 MHz bandwidth. IQ mixers address the problem of maximizing information transmission in a limited bandwidth by allowing the user to modulate both the in-phase and quadrature components of a carrier simultaneously, doubling the information density.

We plan to use a modified version of the ARCONS DSP FPGA firmware. A detailed description of their DSP firmware is found in \cite{McHugh2012}. Due to the increased logic and bandwidth of the ROACH-2 over the previous hardware version, we are able to naively increase the size of DSP processes by a factor of four to scale up to $\sim$2000 resonances per system. The only other major modification to the ARCONS firmware is the replacement of their pulse detection logic with another layer of Digitial Down Conversion (DDC) and we send out I and Q for each MKID at 200Hz continuously.

We baseline the use of four ROACH-2 boards, one for each the 500 and 350 µm arrays, and two for the 250 µm array. The warm readout electronics will be housed in a custom designed box that will provide EM shielding as well as heat sinking elements on the boards to the inner- frame to avoid overheating of the components. The ROACH-2 electronics will be mounted directly to the inner frame near the cryostat to minimize the path length of the readout SMA cables.
\subsection{Polarization Modulation}
\label{ssec:hwpr}

\begin{figure}[t]
  \centerline{
    \includegraphics[height=3in]{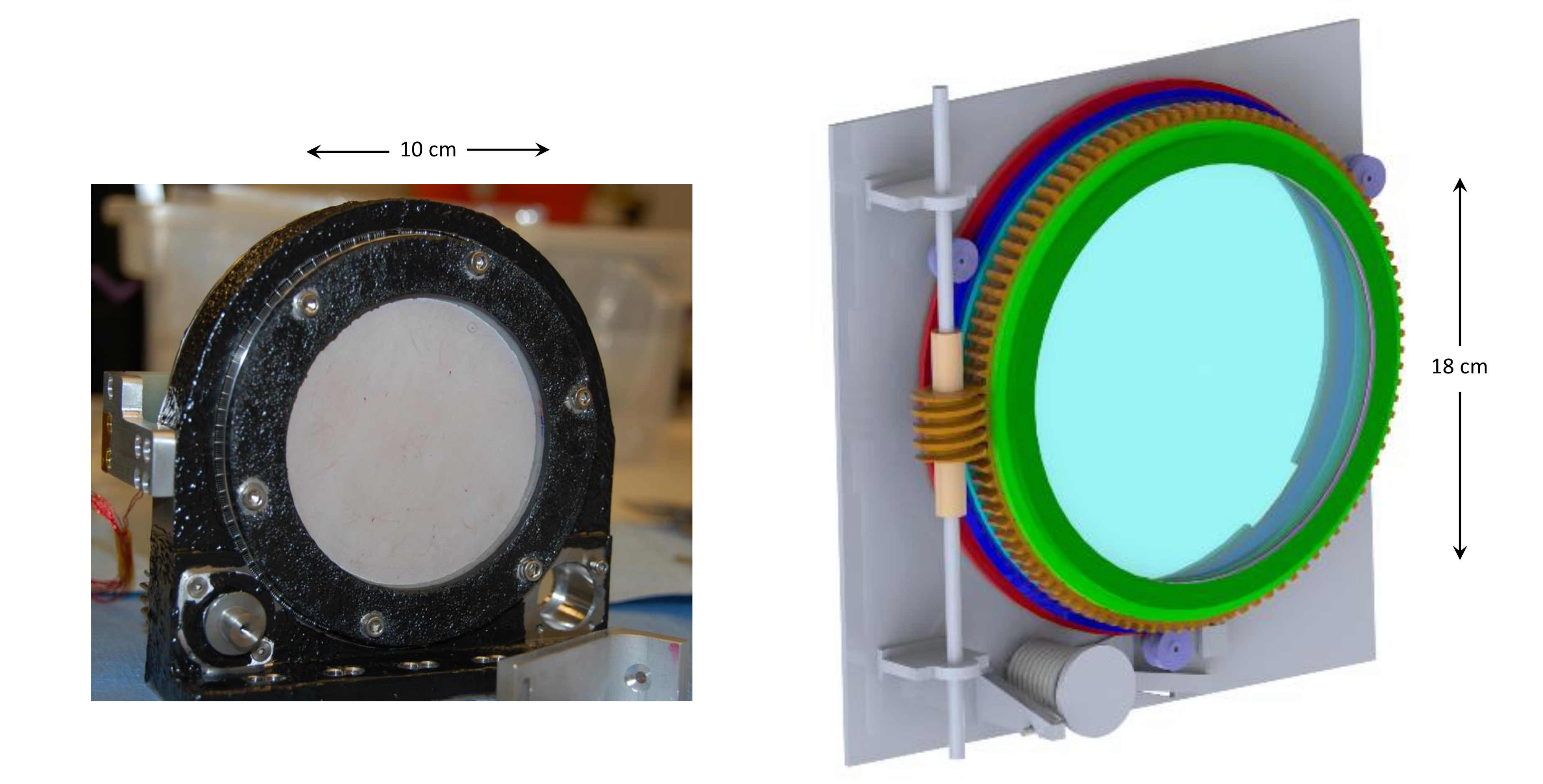}
  }
  \caption{\label{fig:hwpr} Left: An image of the half wave plate and rotator mechanism that was used in BLASTPol. Right: A conceptual drawing of the new wave plate and bearing with a larger clear aperture. There are inevitably imperfections of the AHWP in the manufacturing process that are observable as level shifts in the signal at different AHWP positions. To mitigate this we try to place the AHWP as far from the Cassegrain focus as possible so that the beams from different pixels overlap as much as possible on the waveplate. This minimizes the signal step but drives a larger aperture for the AHWP.}
\end{figure}

Polarimetry is achieved through the dual-polarization-sensitive pixels and a stepped AHWP. The AHWP is turned between four predefined positions (0$^\circ$, 22.5$^\circ$, 45$^\circ$, and 67.5$^\circ$) using a vacuum stepper motor, mounted to the lid of the cryostat, and an absolute encoder, mounted on the rim of the AHWP rotator mechanism. The AHWP is stepped after each full scan of a target and each target is observed at all of the HWP positions. This enables each detector to sample $\pm$Q and $\pm$U multiple times during the mapping of a target. The AHWP used on BLASTPol was constructed using five layers of 500\micron\ thick sapphire, which are glued together by 6\micron\ thick layers of polyethylene (Figure \ref{fig:hwpr}). A broadband anti-reflective coating (ARC), using an artificial dielectric, is placed on the front and back surfaces of the AHWP. The effectiveness of the AHWP across the BLASTPol bands was determined by measuring the nine elements of the linear polarization Mueller matrix \cite{Moncelsi2014}. The diagonal elements of the matrix measure the optical and polarization efficiency and are indicative of the high quality of the AHWP. The band-averaged optical efficiency at 350 and 500 \micron\ is $\sim$$1.0$ and at 250 \micron\ it is $\sim$$0.9$ . The polarization efficiency is $>95\%$ at 350 and 500 \micron\ while it is $\sim$$80 \%$ for the 250 \micron\ channel. The reduced performance at the shortest wavelengths is a result of the exceptionally wide bandwidth that was optimized for the 500 \micron\ band. Details of the construction and testing of the AHWP and ARC can be found in \citealt{Moncelsi2014}.

BLAST-TNG requires a larger aperture AHWP which has motivated the development of alternatives to the sapphire design. One such solution could be a metal-mesh AHWP. These devices are manufactured using the same technology used to produce metal-mesh filters, which are widely used in submillimeter instrumentation \cite{Ade2006}. Both air-gap \cite{Pisano2008} and polypropylene embedded metal-mesh AHWPs \cite{Pisano2012} have been realized at millimeter wavelengths and their feasibility at THz frequencies is under study. Embedded metal-mesh AHWPs can also be produced with diameters larger than the commercially available sapphire plates (i.e. larger than $\sim$33 cm).

\section{Conclusion}
The BLAST-TNG telescope will be a powerful demonstrator of a number of new technologies, especially large diameter carbon fiber mirrors and MKID detector arrays. It will offer an unprecedented look at the structure of magnetic fields in the Galaxy and will serve to constrain models of star formation processes, dust grains, and polarization foregrounds. The instrument is in an advanced stage of development and will see most of the major components manufactured within the next year, followed by extensive integration and testing, and finally, a flight from Antarctica in December 2016.

\section{Acknowledgements}

The BLAST-TNG collaboration acknowledges the support of NASA through grant numbers NNX13AE50G S04 and NNX09AB98G, the Leverhulme Trust through the Research Project Grant F/00 407/BN, Canada's Natural Sciences and Engineering Research Council (NSERC), and the National Science Foundation Office of Polar Programs. The BLAST-TNG collaboration would also like to acknowledge the Xilinx University Program for their generous donation of 4 Virtex-6 FPGAs for use in our ROACH-2 readout electronics. Frederick Poidevin thanks the Spanish Ministry of Economy and Competitiveness (MINECO) under the Consolider-Ingenio project CSD2010-00064 (EPI: Exploring the Physics of Inflation). Bradley Dober is supported through a NASA Earth and Space Science Fellowship. Peter Ashton is supported through Reach for the Stars, a GK-12 program supported by the National Science Foundation under grant DGE-0948017. However, any opinions, findings, conclusions, and/or recommendations are those of the investigators and do not necessarily reflect the views of the Foundation. We would also like to thank the Columbia Scientic Balloon Facility (CSBF) staff for their continued outstanding work.

\bibliography{BLAST-TNG_JAI_v2}
\bibliographystyle{ws-jai} 
\end{document}